\documentclass[prl,twocolumn,superscriptaddress,nobibnotes]{revtex4-1}

\usepackage{graphicx}
\usepackage{dcolumn}
\usepackage{bm}
\usepackage{color}
\newcommand{\tto}[1]{Tb$_2$Ti$_2$O$_7${#1}}

\begin{document}


\title{Magnetoelastic excitations in the pyrochlore spin liquid Tb$_2$Ti$_2$O$_7$}
\author{T. Fennell}
\email{tom.fennell@psi.ch}
\altaffiliation{Began this work at Institut Laue Langevin, BP 156, 6, rue Jules Horowitz, 38042, Grenoble Cedex 9, France}
\affiliation{Laboratory for Neutron Scattering, Paul Scherrer Institut, 5232 Villigen PSI, Switzerland}
\author{M. Kenzelmann}
\affiliation{Laboratory for Developments and Methods, Paul Scherrer Institut, 5232 Villigen PSI, Switzerland}
\author{B. Roessli}
\affiliation{Laboratory for Neutron Scattering, Paul Scherrer Institut, 5232 Villigen PSI, Switzerland}
\author{H. Mutka}
\affiliation{Institut Laue Langevin, BP 156, 6, rue Jules Horowitz, 38042, Grenoble Cedex 9, France}
\author{J. Ollivier}
\affiliation{Institut Laue Langevin, BP 156, 6, rue Jules Horowitz, 38042, Grenoble Cedex 9, France}
\author{M. Ruminy}
\affiliation{Laboratory for Neutron Scattering, Paul Scherrer Institut, 5232 Villigen PSI, Switzerland}
\author{U. Stuhr}
\affiliation{Laboratory for Neutron Scattering, Paul Scherrer Institut, 5232 Villigen PSI, Switzerland}
\author{O. Zaharko}
\affiliation{Laboratory for Neutron Scattering, Paul Scherrer Institut, 5232 Villigen PSI, Switzerland}
\author{L. Bovo}
\affiliation{London Centre for Nanotechnology and Department of Physics and Astronomy, University College London, 17-19 Gordon Street, London, WC1H 0AH, UK}
\author{A. Cervellino}
\affiliation{Swiss Light Source, Paul Scherrer Institut, 5232 Villigen PSI, Switzerland}
\author{M. K. Haas}
\altaffiliation{Now at Air Products and Chemicals Inc., Allentown PA 18195 USA}
\affiliation{Department of Chemistry, Princeton University,  Princeton NJ 08540, USA} 
\author{ R. J. Cava}
\affiliation{Department of Chemistry, Princeton University,  Princeton NJ 08540, USA}

\date{\today}

\begin{abstract}
At low temperatures, \tto{} enters a spin liquid state, despite expectations of magnetic order and/or a structural distortion.  Using neutron scattering, we have discovered that in this spin liquid state an excited crystal field level is coupled to a transverse acoustic phonon, forming a hybrid excitation.  Magnetic and phonon-like branches with identical dispersion relations can be identified, and the hybridization vanishes in the paramagnetic state.   We suggest that \tto{} is aptly named a ``magnetoelastic spin liquid'' and that the hybridization of the excitations suppresses both magnetic ordering and the structural distortion.  The spin liquid phase of \tto{} can now be regarded as a Coulomb phase with propagating bosonic spin excitations. 
\end{abstract}

\maketitle

Spin liquids~\cite{Balents:2010jx} are often defined as correlated but fluctuating spin states with unbroken translation and spin rotation symmetry.  In theory, many types of spin liquid can exist~\cite{Wen:2002cy}, but their experimental identification and classification is problematic.  Since the absence of broken symmetry alone is not definitive and topological properties~\cite{Wen:2002cy} are not experimentally accessible, one possibility is to study their excitations.  These are often predicted to be exotic fractional quasiparticles such as spinons~\cite{Han:2012fo} or monopoles~\cite{Castelnovo:2008hb}, but propagating bosonic excitations are possible in certain models~\cite{Robert:2008ib,Hermele:2004p9,Benton:2012ep,Levin:2005th}. 

\tto{}, which is often referred to as a spin liquid, does indeed remain in a magnetically disordered phase with spin dynamics down to 0.05 K~\cite{Gardner:2010fu}.  The Tb$^{3+}$ ions form a pyrochlore lattice and the spin interactions are antiferromagnetic ($\theta_{CW}=-19$ K), but the crystal field splits the $^7F_6$ free ion term of Tb$^{3+}$ to give a doublet groundstate with Ising character.  Classically, such a spin system should order, with $T_N\sim 1-2$ K predicted for \tto{}~\cite{denHertog:2000tc,Kao:2003ex}.  Instead,  the spin liquid state of \tto{} develops below $T\sim20$ K.  At low temperature, pinch points appear in the diffuse neutron scattering, suggesting that this is a magnetic Coulomb phase governed by ice rules~\cite{Fennell:2012ci,Petit:2012ko,fritsch,Guitteny:2013hf}.  

However, because Tb$^{3+}$ is a non-Kramers ion, its degenerate electronic states are susceptible to Jahn-Teller distortions~\cite{Gehring:1975uk}.  There is much experimental evidence of magnetoelastic effects below $T\sim 20$ K - the Young's modulus and elastic constants soften very significantly~\cite{MAMSUROVA:1986wx,*MAMSUROVA:1988wg,Nakanishi:2011bz}; structural Bragg peaks broaden anisotropically~\cite{Ruff:2007hf}; there is a large dielectric anomaly~\cite{Mamsurova:1985ug}; the low temperature state is susceptible to pressure-induced magnetic order~\cite{Mirebeau:2002fm} and magnetic field-induced structural modifications~\cite{Ruff:2010uq}; acoustic phonons are strongly scattered by the spins~\cite{Li:2013ko} - but no distortion has been observed.  

The strong expectations of long-range order and/or a structural distortion mean that the true nature of the spin liquid state and the mechanism of its existence are not evident.  Usually, magnetoelastic coupling is a mechanism for the relief of frustration~\cite{Tchernyshyov:2002ih,Chern:2006us}, but the spin liquid state in \tto{} exists throughout the same temperature regime as the anomalous elastic properties, leading to the suggestion that both spin and structural degrees of freedom are frustrated in \tto{}~\cite{taniguchi}.  

No theory simultaneously accounts for both the magnetoelastic phenomena and the spin liquid.  Models based on single-ion magnetostriction mechanisms reproduce the bulk magnetoelastic properties~\cite{MAMSUROVA:1986wx,*MAMSUROVA:1988wg,Klekovkina:2011vv} but make no account of the spin liquid; theories which focus on the evasion of long-range magnetic order by the introduction of quantum fluctuations by virtual crystal field excitations~\cite{Molavian:2007ig}, hypothetical distortions~\cite{Curnoe:2008gy,Bonville:2011dw}, or anisotropic exchange~\cite{Curnoe:2013iz} are successful in  reproducing features of the diffuse neutron scattering~\cite{Molavian:2007ig,Curnoe:2008gy,Petit:2012ko,Curnoe:2013iz}, but have individual drawbacks (a magnetization plateau predicted in the case of virtual crystal field excitations~\cite{Molavian:2009dl} is strongly debated~\cite{legl2012,*Baker:2012cd,*Lhotel:2012hb,*Yin:2013iz,mirebeauarxiv}; distortions~\cite{Curnoe:2008gy,Bonville:2011dw} remain hypothetical, and single-ion singlet groundstates~\cite{Bonville:2011dw} cannot account for the large elastic magnetic spectral weight~\cite{Gaulin:2011ba}) and make no explanation of the magnetoelastic behaviour.

\begin{figure*}
\includegraphics[scale=0.6,trim=50 170 90 80,angle=0]{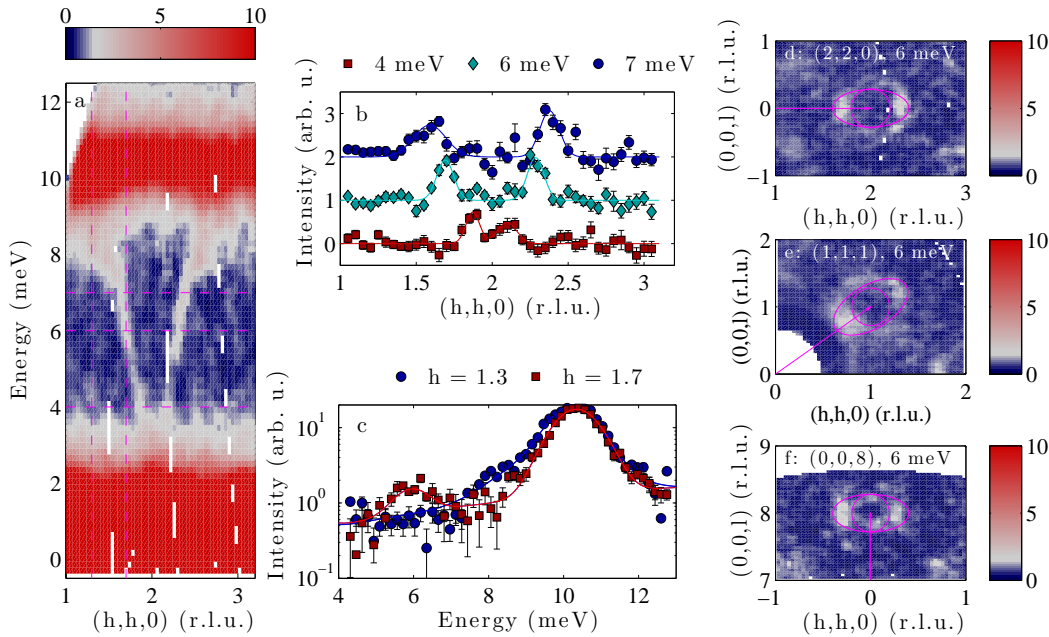}
\caption{Overview of the MEM.    The sharp mode between the two intense  crystal field excitations at 1.5 and 10 meV is the MEM (a).  It is weak, but can be clearly identified in constant-energy cuts (b, background levels offset by 1) and constant-$\mathbf{Q}$ cuts (c) (cut positions are indicated on (a) by dashed lines).  Constant-energy maps show the MEM at $(2,2,0)$ (d) and a similar mode at $(1,1,1)$ (e).  At these small wavevectors, the modes have their intensity parallel to the scattering vector (the arrow and ellipsoids show the scattering vector and highlight the intensity distribution respectively).  Excitations with similar dispersions are visible at large wavevectors (e.g. $(0,0,8)$, f), but with a transverse intensity distribution.}
\label{f1}
\end{figure*}

We contend that the electronic and structural excitations of \tto{} are mixed into hybrid fluctuations which we call magnetoelastic modes (MEMs), and this is at the origin of the absence of magnetic order and structural distortion in \tto{}.  We characterize a MEM, and demonstrate that it has both magnetic and phononic characters, which are visible at different wavevectors.

We used the same single crystal of \tto{} as in Ref.~\cite{Fennell:2012ci}.  It has no sign of any ordering transition between 0.3 and 50 K in its heat capacity, and by comparison with the series of Tb$_{2+x}$Ti$_{2-x}$O$_{7-x/2}$ powders reported in~\cite{taniguchi}, its lattice parameter ($a=10.155288(1)$ \AA) suggests its stoichiometry is Tb$_{2.013\pm0.002}$Ti$_{1.987\pm0.002}$O$_{6.994\pm0.001}$.  Further details of its characterization are to be found in the supplementary material~\cite{supmat}.  Using the time-of-flight (TOF) spectrometer IN5~\cite{Ollivier:2010wu} at the Institut Laue Langevin, we surveyed a four-dimensional volume of $S(\mathbf{Q},\omega)$.  We measured at 0.05, 5 and 20 K using $\lambda_i=4$ \AA, and additionally at 0.05 K using $\lambda_i=2$ and $7$ \AA.  Using the triple-axis spectrometers (TAS) TASP (in combination with the neutron polarimetry device MuPAD) and EIGER  at the SINQ, Paul Scherrer Institut, we investigated the polarization and temperature dependence of the MEM respectively.  In the polarized neutron scattering experiment, we measured non-spin flip (NSF) and spin flip (SF) cross sections with the neutron polarization parallel to the scattering vector $\mathbf{Q}$, such that all magnetic scattering appears in the SF channel and nuclear scattering in the NSF channel.  Data from the TAS experiments can be compared with cuts through the TOF spectra by scaling.

\begin{figure}
\includegraphics[scale=0.6,trim=140 50 350 40,angle=0]{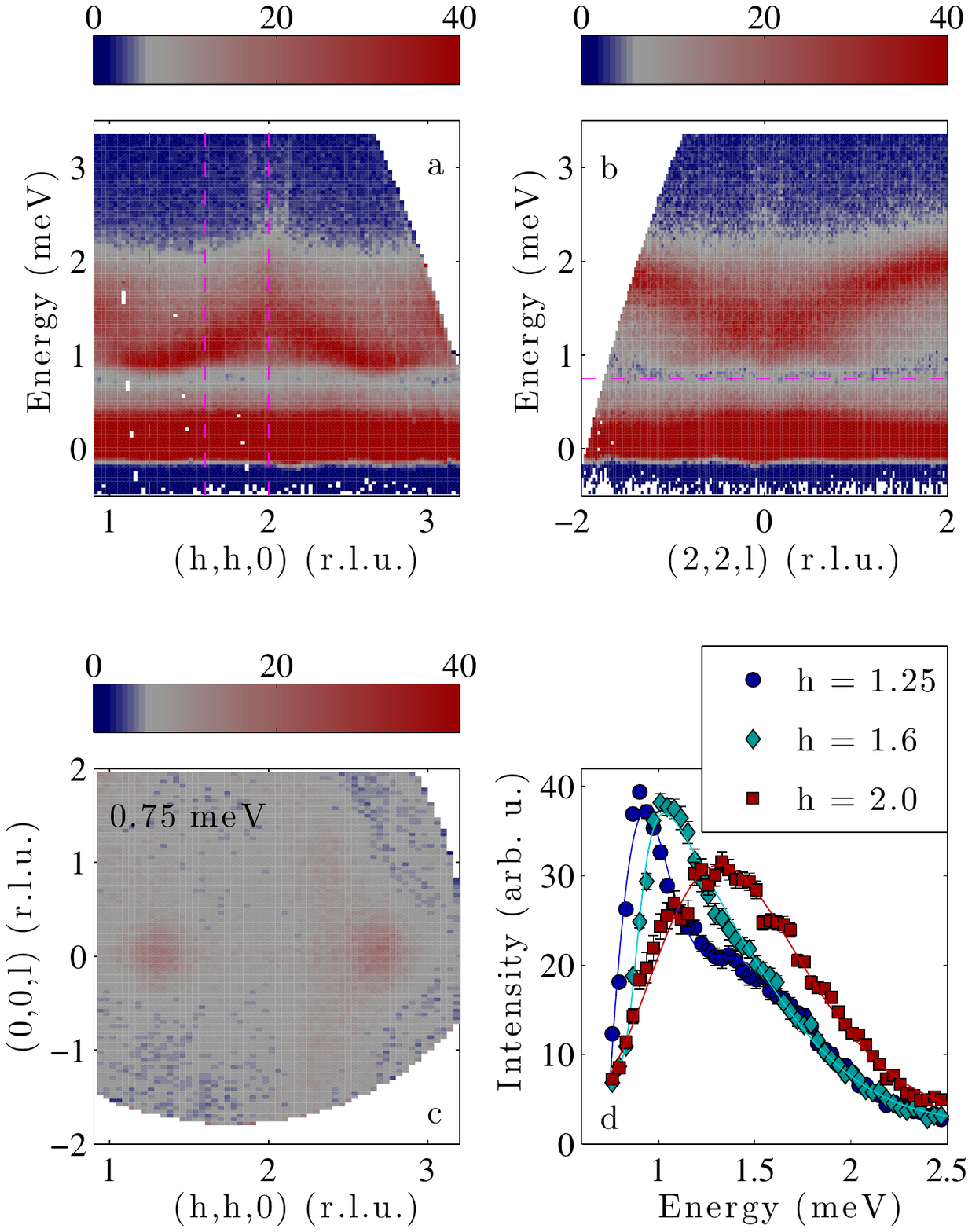}
\caption{Interaction of the MEM with the first crystal field excitation.    At $(2,2,0)$, the MEM intersects with the firstCFE (a, this is the same view as Fig.~\ref{f1}a, but with better energy resolution).  The MEM can be seen above the CFE.  The MEM is faintly visible in the perpendicular direction (b), where some intensity concentrated around $(h,h,0)$ is integrated in the cut.  No sharp feature can be found extending below the CFE, as can also be seen in an intensity map at $E=0.75$ meV (dashed line in b), i.e. in the gap (c).  Constant-$\mathbf{Q}$ cuts (d, at positions indicated by dashed lines in panel a) show a two-component lineshape for the CFE far away from $(2,2,0)$ and a single, broad, asymmetric, peak where the MEM meets the crystal field level. }
\label{f2}
\end{figure}

The known magnetic neutron scattering response of \tto{} consists of elastic diffuse scattering~\cite{Gardner:2001kv,Fennell:2012ci,fritsch}, quasielastic scattering~\cite{Petit:2012ko,Yasui:2002p6660}, and crystal field excitations (CFEs)~\cite{Gardner:2010fu}.  Fig.~\ref{f1} and~\ref{f2} show overviews of the inelastic scattering, with lower resolution extending to higher energy transfer, and with higher energy resolution around the first CFE, respectively.  We concentrate here on the MEM, which is a new feature.  It is the weak but sharp mode extending out of the $(2,2,0)$ position, between the two intense CFEs (Fig.~\ref{f1}a-d). A similar mode is visible at $(1,1,1)$ (Fig.~\ref{f1}e), and the topmost part of the dispersion can be distinguished in nearby zones.  Strong excitations are also visible at $(0,0,8)$ (Fig.~\ref{f1}f), $(3,3,7)$ and $(5,5,5)$, but while those in low zones have their propagation vector ($\mathbf{k}$) parallel to the scattering vector ($\mathbf{Q}$), these have $\mathbf{k}\perp\mathbf{Q}$.  

The first CFE is itself quite significantly dispersive at low temperatures.  The interaction of the MEM with the first CFE, which is pulled up in energy where the two meet, can be seen in Fig.~\ref{f2}a.  There is no branch of the MEM reaching down to $\hbar\omega=0$, below the CFE (Fig. ~\ref{f2}a, b, c).  Examination of the intensity throughout the iso-energy volume $S(\mathbf{Q},\hbar\omega=0.65~\mathrm{meV})$ shows no sharp features cut through it (Fig.~\ref{f2}c).  A broad, asymmetric peak is formed where the modes intersect, but, away from $(2,2,0)$, two components can be distinguished in the CFE (Fig.~\ref{f2}d).   The MEM does not  interact with the second CFE, as its dispersion passes just below it (Fig.~\ref{f1}c).

\begin{figure}
\begin{center}
\includegraphics[scale=0.6,trim=110 110 120 240,angle=0]{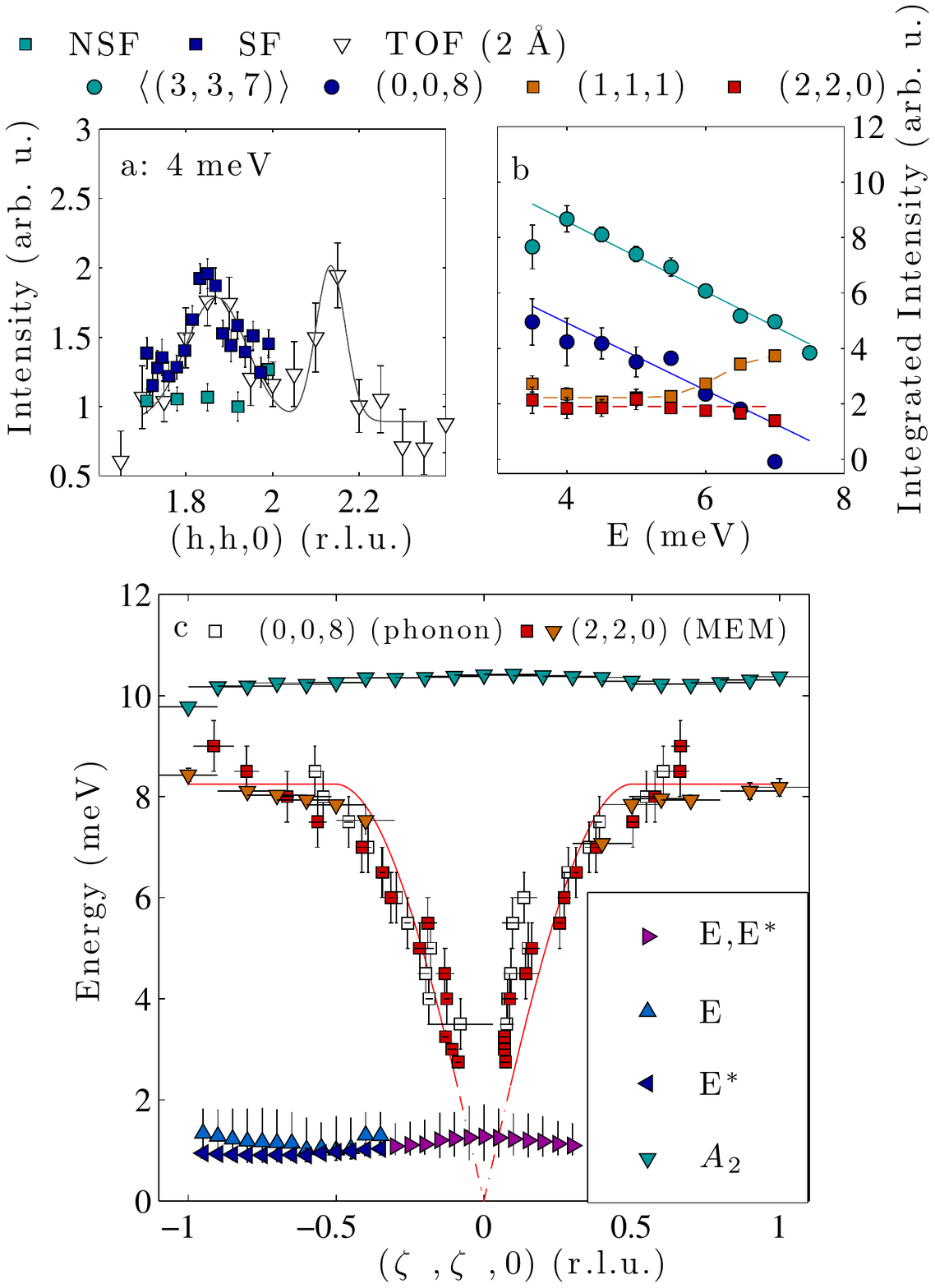}
\end{center}
\caption{Characterization of the excitations in \tto{} at 0.05 K.  Polarization analysis (a) shows that the scattering observed in the MEM at $(2,2,0)$ is magnetic.  The integrated intensities (b) show that the modes at large wavevector have intensities which decrease with $\hbar\omega$, similar to phonons, while those at $(2,2,0)$ and $(1,1,1)$ have a different character.  Fitted peak positions from constant-energy cuts (squares) and constant-$\mathbf{Q}$ cuts (triangles) show that the dispersion of the magnetic mode is exactly the same as the phonon-like mode (c).  $A_2$ is the second CFE, $E$ and $E^*$ are the two components of the first CFE, where they can be distinguished.  This level has been fitted with asymmetric Gaussian functions, and the bar indicates the half-maximum height of the asymmetric lineshapes.  Elsewhere, the two bars indicate the error of the fitted peak position and the width of the integral used for the cut.  Integrated intensities were extracted from energy slices such as Fig. 1d by summing all of the intensity above background within a ring fixed by the dispersion.}
\label{f3}
\end{figure}

MEMs can be observed at small wavevectors, typical of magnetic excitations.  At $(2,2,0)$, we have determined explicitly that the MEM has a magnetic contribution by using polarized neutron scattering.  As shown in Fig.~\ref{f3}a, all the scattering occurs in the spin flip channel, indicating that in fact there is no measurable nuclear contribution at this position.  The intensity of the MEM is almost independent of energy until it approaches the second CFE where it may increase (Fig.~\ref{f3}b), in contrast to typical antiferromagnetic excitations which decrease in intensity with increasing energy.  

Because magnetic neutron scattering is due to spin components perpendicular to $\mathbf{Q}$, and the wavevector $\mathbf{k}$ of the mode is parallel to ${\mathbf{Q}}$, the MEM is a transverse mode (i.e. the spin fluctuations are perpendicular to its direction of propagation).   In comparison, the excitations at large wavevectors are similar to transverse acoustic phonons - they appear at large wavevectors  (the phonon cross section depends on $|\mathbf{Q}|^2$), they are gapless (within the energy resolution of this setting of the spectrometer), their intensity decreases with $\hbar\omega$ (Fig.~\ref{f3}b), and they are intense when $\mathbf{k}\perp\mathbf{Q}$.  However, if we compare the transverse magnetic mode at $(2,2,0)$ with the transverse phonon-like mode $(0,0,8)$ (which is to say along $(h,h,0)$ and $(h,h,8)$ respectively), we see that the upper parts of their dispersions overlap precisely (Fig.~\ref{f3}c), suggesting they have a common origin, and that these are therefore mixed modes carrying both magnetic and structural fluctuations.  We observe the magnetic part at small wavevectors, where the magnetic form factor of Tb$^{3+}$ is large, and the phononic part at large wavevectors, where the phonon cross section is large.

\begin{figure}
\begin{center}
\includegraphics[scale=0.5,trim=210 320 130 300,angle=0]{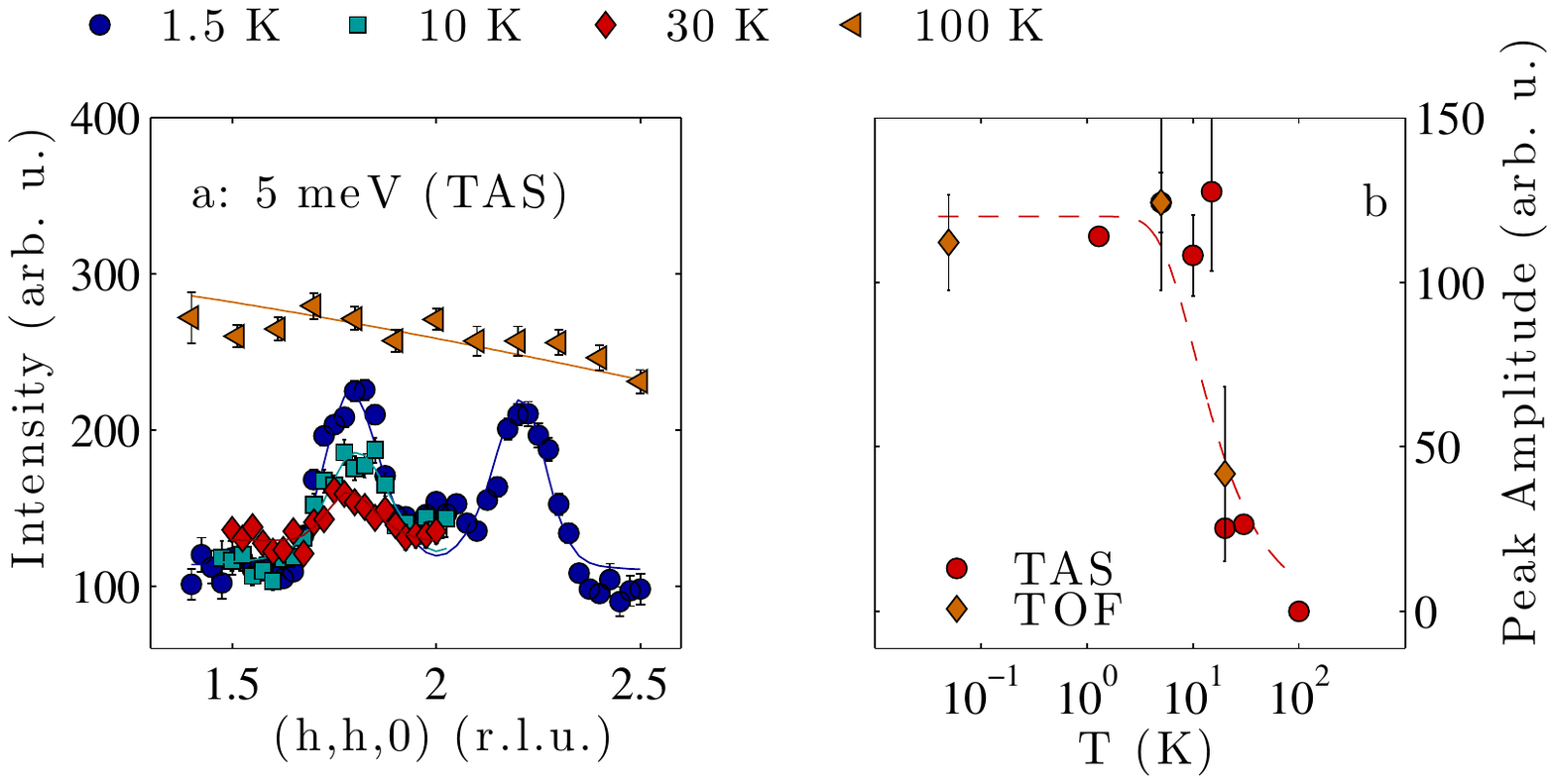}
\end{center}
\caption{Temperature dependence of the MEM.  The MEM intensity collapses above 10 K (a).  As the peak intensity falls above 10 K (b), a rising intensity which follows the magnetic form factor of Tb$^{3+}$ (a, 100 K) also appears.  This contribution originates from the thermal broadening of the nearby CFEs (it follows the Bose factor).  The dashed line in b is $n_0-n_1$ (scaled), where $n_0$ and $n_1$ are the thermal population factors of the ground and first excited states respectively of a two-level system with $\Delta=1.4$ meV.  TOF peak areas scaled to TAS peak amplitudes at 5 K.}
\label{f4}
\end{figure}

In Fig.~\ref{f4}, we show the temperature dependence of the MEM at $(2,2,0)$.  Its intensity collapses in the range $10-20$ K, again quite unlike a conventional phonon, which would become stronger at higher temperature.  The detailed structure of  interacting modes and two branches of the first CFE shown in Fig.~\ref{f2} also collapses at $T\sim 20$ K.  The CFE becomes a single level with (almost) none of the dispersion visible in Fig.~\ref{f2}a.   

The low temperature dispersion of the first CFE indicates that the high-temperature single-ion excitations have been replaced by propagating excitons.  The appearance of two branches in the exciton band indicates that, as in other rare earth pyrochlores, the exchange is anisotropic (different fluctuation directions become non-degenerate).  The dispersion of these branches will provide a means to determine the anisotropic exchange parameters in \tto{}~\cite{Houmann:1979ei}.

In the usual crystal field scheme of \tto{}, the ground and first excited states (dominantly  $a_4|\pm4\rangle\pm a_5|\mp5\rangle$ and $\pm b_5|\pm5\rangle+ b_4|\mp4\rangle$ respectively) are connected not only by the operators $J_\pm$, but also $J_x$, $J_y$, and the quadrupole operators $O_{xz}=J_x J_z + J_z J_x=1/2[(J_+ J_z + J_z J_+)+(J_- J_z + J_z J_-)]$ and  $O_{yz}=J_y J_z + J_z J_y=1/(2i)[(J_+ J_z + J_z J_+)-(J_- J_z + J_z J_-)]$.  The finite matrix elements of $J_x$ and $J_y$ mean that the excitons are transverse fluctuations, and because of the quadrupole operators, they can mix with the transverse phonons~\footnote{This argument is correct for fluctuations propagating along $[111]$, the quantization direction of $J_z$ for a single sublattice.  Some departure from this description in terms of purely transverse fluctuations may be expected for the spins or fluctuations in other directions.}.  The general features of such a coupling, which is linear in the relevant operators, are that its strength is largest at high energy, but decreases as the energy difference between the modes increases; it vanishes as $(\mathbf{k},\hbar\omega)\to0$ and as the population of the CFE becomes comparable to that of the groundstate~\cite{Jensen:1991ux}.  This is in qualitative agreement with our observations - the MEM is undetectable at low energy (Fig.~\ref{f2}), while competition between the first two factors may result in the intensity distribution in the MEM shown in Fig.~\ref{f3}b.  The MEM vanishes in the relevant temperature interval, as shown in Fig.~\ref{f4}.  This temperature scale is also that in which the spin correlations evolve most strongly~\cite{Fennell:2012ci}, suggesting that the magnetoelastic effects are not a coincidental property of the spin liquid phase. 

The derivation of a Hamiltonian for \tto{} remains challenging.   The measurement of the excitation spectrum throughout a large volume of $S(\mathbf{Q},\hbar\omega)$ shows no indication of global symmetry lowering or a soft mode associated with a structural transition.  The key to the evasion of long-range magnetic order in \tto{} seems to be the mixing of the first crystal field level with the groundstate, which has been attempted theoretically in different ways~\cite{Kao:2003ex,Curnoe:2007hd,Molavian:2007ig,Bonville:2011dw,Curnoe:2013iz}, while the coupling of excitons and phonons we have observed suggests that both the quadrupole operators and anisotropic exchange are of central importance.  

The low temperature state of \tto{} is ever more intriguing.  In the spin sector we may hope for an emergent gauge theory, which must now contain power-law spin correlations~\cite{Fennell:2012ci} and a propagating bosonic excitation.  Various theories of frustrated magnetism support dispersive excitations despite the absence of long-range magnetic order.  In a quantum spin ice, the photon mode~\cite{Hermele:2004p9,Benton:2012ep} looks superficially much like the MEM, and we speculate that the theory of a magnetoelastic spin liquid will ultimately resemble a quantum spin ice, with vibronic fluctuations replacing the quantum tunneling fluctuations.  In this context, the microscopic meaning of our results is that exchange interactions and atomic wavefunctions depend on the position of atoms, which can themselves fluctuate in a correlated manner.  Since an acoustic phonon is involved, passage of a hybrid fluctuation can reconfigure both spins and wavefunctions over a large distance.  We suggest that \tto{} should be viewed as an example of dynamical frustration~\cite{Molavian:2007ig,goremychkin} mediated by the spin-lattice coupling~\cite{taniguchi} evidenced here.

In conclusion, we have observed a magnetoelastic mode in the spin liquid phase of \tto{}.  This mode is formed by the hybridization of the first excited crystal field level and the transverse acoustic phonons.  The hybridization of the excitations disappears in the paramagnetic regime.  We suggest that the coupling we have observed is at the origin of the anomalous magnetoelastic behavior of \tto{}.  The existence of the magnetoelastic mode implies that the spin liquid phase of \tto{} is a Coulomb phase supporting a propagating bosonic spin excitation.  

{\it Note added} - Since the submission of this paper measurements of the quasielastic scattering have shown that it also contains a propagating mode~\cite{Guitteny:2013hf}.  The MEM at $(1,1,1)$ is clearly visible in this study, but was interpreted as an acoustic phonon.

\begin{acknowledgements}
We thank X Thonon (ILL), M Zolliker and M Bartkowiak (PSI) for operation of dilution refrigerators; S Fischer and W Latscha for additional cryogenic support on EIGER; and A Bullemer for assistance with cutting the sample.  We are pleased to acknowledge discussions with O Benton, AT Boothroyd, ST Bramwell, MJP Gingras, A Gukasov, H Kadowaki, PA McClarty, C Ruegg, N Shannon, and particularly J Jensen. Neutron scattering experiments were carried out at the high flux reactor of the Institut Laue Langevin in Grenoble, France, and the continuous spallation neutron source SINQ at the Paul Scherrer Institut at Villigen PSI in Switzerland, and work at PSI was partly funded by the Swiss National Science Foundation grant ``Quantum Frustration in Model Magnets'' (200021\_140862).
\end{acknowledgements}


%

\begin{center}
{\bf Supplementary Information}
\end{center}
In Ref.~\cite{taniguchi} it was suggested that \tto{} samples are of variable quality, having only the approximate stoichiometry Tb$_{2+x}$Ti$_{2-x}$O$_y$, and that different properties vis-a-vis the spin liquid state or magnetic order result.  A similar conclusion about samples of Yb$_2$Ti$_2$O$_7$ was drawn in Ref.~\cite{Ross:2012dj}.  Because of the recently highlighted sample dependence we present some further characterizations of our sample by synchrotron powder x-ray diffraction and heat capacity measurements, and also refer briefly to a single crystal neutron diffraction experiment.  We suggest that the results will provide a guide as to which \tto{} samples would be expected to show a MEM and associated magnetoelastic spin liquid, as compared to the ordered groundstate found in Ref.~\cite{taniguchi}.  The following information contains: {\it i: Background}; {\it ii: Sample details}; {\it iii: Heat capacity measurements}; {\it iv: X-ray diffraction measurements}; {\it v: Discussion and Summary}.

{\it i: Background.} Long-ranged magnetic order, and/or a splitting of the groundstate doublet into two singlets are highly plausible in \tto{}, but have only been observed in certain powder samples, depending on very fine control of the stoichiometry~\cite{taniguchi}.  In that work, a well-developed peak appears in the heat capacity at $T\sim0.5$ K, and inelastic neutron scattering clearly shows a new crystal field excitation at 0.1 meV.  In other works, the sample (which may be a powder or a crystal) shows a poorly developed peak in the heat capacity, or no peak at all~\cite{Siddharthan:1999ww,Gingras:2000vd,Hamaguchi:2004dt,Cornelius:2005tg,Ke:2009ds,Chapuis:2010ir,Yaouanc:2011be,Takatsu:2011kg}.  Furthermore, there are variable reports of spin freezing at $T\sim0.4$ K observed by ac-susceptibility, $\mu$SR or neutron spin echo measurements~\cite{Gardner:2010fu}.   In Ref.~\cite{taniguchi} it was suggested that \tto{} samples are of variable quality, having only the approximate stoichiometry Tb$_{2+x}$Ti$_{2-x}$O$_y$.  A spin liquid state was attributed to samples with $x<x_c= -0.0025$, and long range order was found for larger values of $x$.     

Despite this sample dependence, all neutron scattering studies of \tto{} in the literature which use large single crystals are in agreement where different studies and spectrometers can be compared.  Apart from resolution effects, energy cuts through the quasielastic scattering of single crystals appear identical in Refs.~\cite{Petit:2012ko,Yasui:2002p6660} and~\cite{Gaulin:2011ba}, and our high resolution data is also closely comparable.  Intensity maps of the quasielastic scattering in Ref.~\cite{Petit:2012ko} are identical to similar cuts through our high resolution data at casual levels of comparison.  Diffuse scattering at high temperature~\cite{Gardner:2001kv} and low temperature ~\cite{Petit:2012ko,Yasui:2002p6660,fritsch}, appears identical to that which we have measured in our previous work~\cite{Fennell:2012ci}, when the use of polarization analysis is taken into account.  The form of the first crystal field excitation along $(0,0,l)$ reported by Rule {\it et al.}~\cite{Rule:2006fr} and again by Gaulin {\it et al.}~\cite{Gaulin:2011ba} also appears identical to that measured in our sample.  

We therefore suppose that the magnetoelastic spin liquid behaviour we have described is a generic property of ``normal'' samples of \tto{} (i.e. those which have previously been described as spin liquids).  We suggest that the results presented below will provide a guide as to which \tto{} samples would be expected to show a MEM and associated magnetoelastic spin liquid, as compared to the ordered groundstate found in Ref.~\cite{taniguchi}.

\renewcommand{\figurename}{Supporting Information Fig.}
\setcounter{figure}{0}

{\it ii: Sample details.} Our sample is a large single crystal ($\sim7$ g) grown by the floating zone method.  Post-growth, the crystal was mainly dark red in colour, with some black regions.  Annealing under argon converted the crystal completely to the dark red colour.  We cut small pieces from one end of the crystal and used them for specific heat, synchrotron powder x-ray diffraction and single crystal neutron diffraction experiments.  Although these pieces do not come from the very heart of the boule, they do come from the same part which we use for our neutron scattering experiments.  Furthermore, although we are using very small pieces to characterize a large crystal, experiments on the full crystal volume (for example our time of flight spectroscopy, or the neutron Laue diffraction used for alignment of the sample) do not reveal any multiple Bragg peaks which could be suggestive of serious inhomogeneity. 

\begin{figure}
\includegraphics[scale=0.48,trim=100 220 100 230,angle=0]{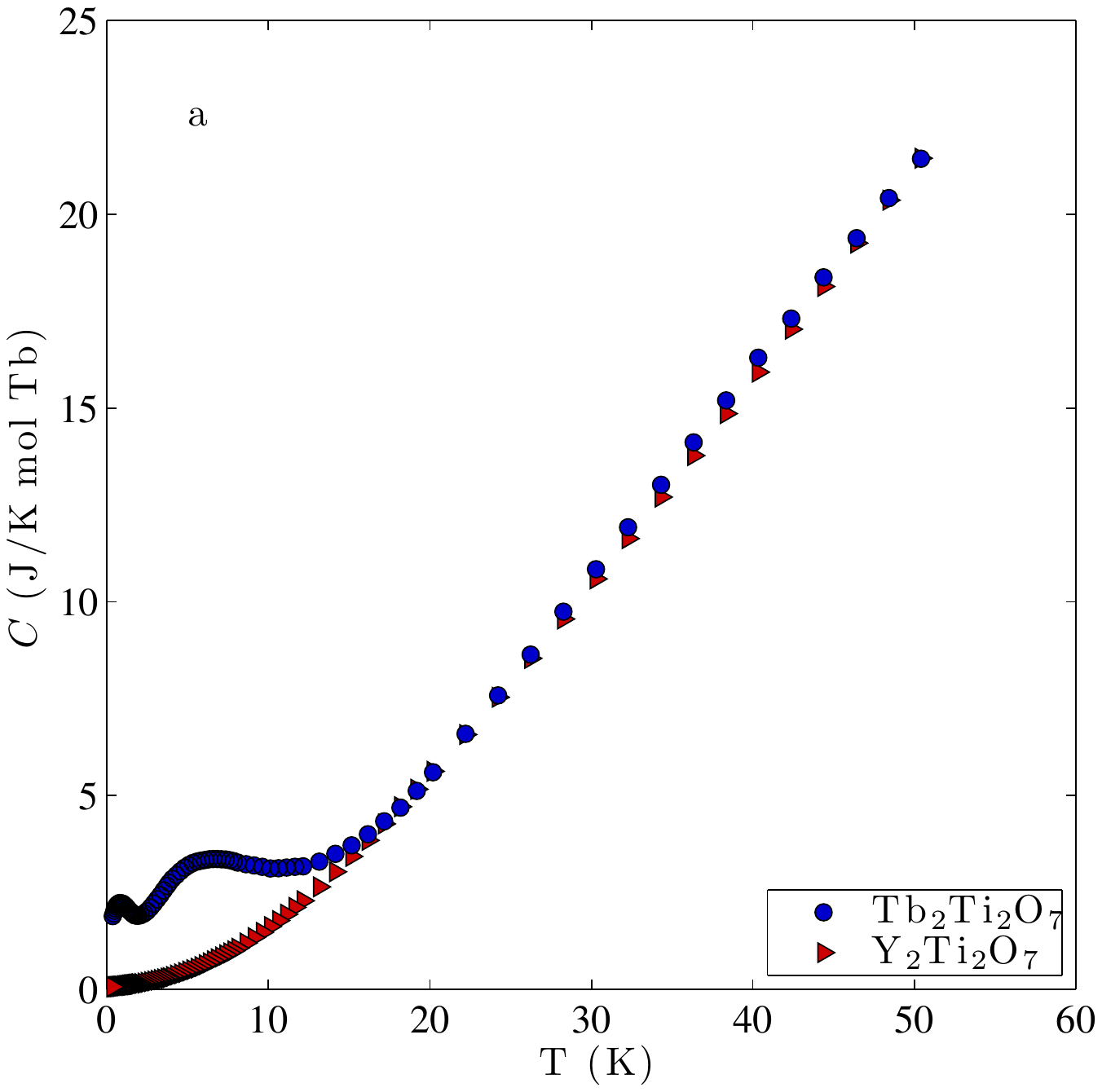}
\includegraphics[scale=0.48,trim=100 220 100 230,angle=0]{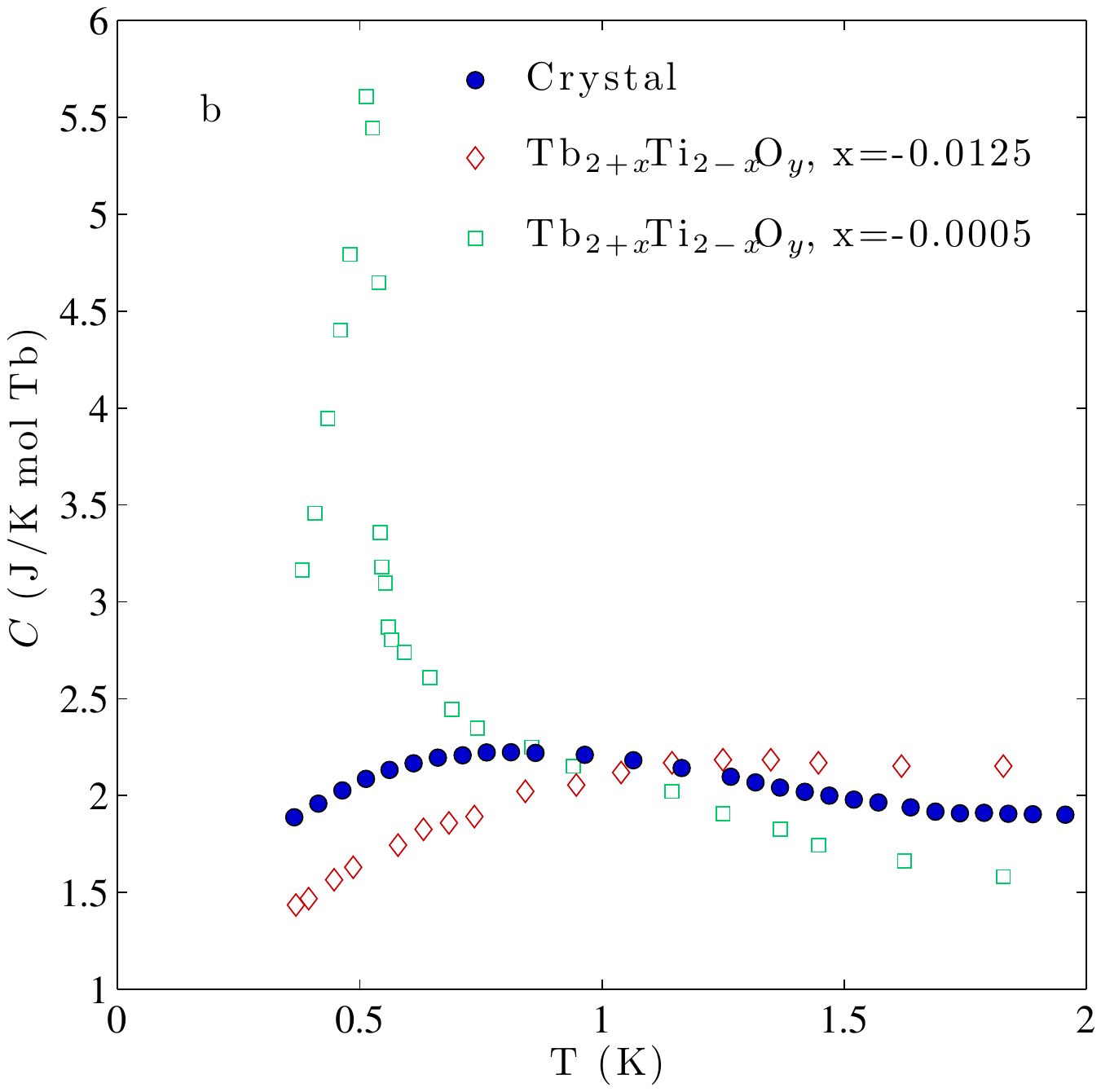}
\caption{Heat capacity of the \tto{} single crystal.  Throughout the full measured temperature range there is no sharp feature associated with a phase transition (a, compared to Y$_2$Ti$_2$O$_7$).   In particular, the low temperature part of the specific heat (b) of our \tto{} sample shows no clear upturn or any development of the peak which is associated with magnetic ordering in certain Tb$_{2+x}$Ti$_{2-x}$O$_y$ samples~\cite{taniguchi}.}
\label{heatcapfig}
\end{figure}

{\it iii: Heat Capacity.} The specific heat of a small piece of the crystal was measured between 0.35 K and 50 K with a Quantum Design Physical Properties Measurement System (PPMS), equipped with a $^3$He option, using a heat-relaxation method. An addenda measurement was made to evaluate the background of Apiezon Grease N and this contribution was subtracted from the data.  Different pyrochlores present different lattice contributions, making an accurate subtraction difficult to evaluate at relatively high temperature, above 10 K, which would be needed to properly estimate the presence of a tail in the magnetic specific heat extending above 10 K.  Nonetheless, we can see that the heat capacity of our sample, reported in Supporting Fig.~\ref{heatcapfig}, most closely resembles that of Refs.~\cite{Ke:2009ds} or~\cite{Yaouanc:2011be}. No sharp features are present in the low temperature regime, below 0.5 K, so the heat capacity presents the form typically associated with the development of the spin liquid state in \tto{}.  The strongest contrast is with the samples of specific stoichiometry discussed by Taniguchi {\it et al}~\cite{taniguchi}, where a sharp peak develops at 0.5 K, and is thought to be associated with magnetic order and splitting of the crystal field doublet states.  The heat capacity data also shows that  there is no sign of a phase transition associated with the onset of the magnetoelastic coupling ($20-30$ K), supporting our suggestion that the magnetoelastic spin liquid develops without global symmetry breaking, and is stable at all temperatures below $T \sim~ 20$ K in normal samples of \tto{}. 

{\it iv: Powder X-ray diffraction.} Two tiny fragments from the crystal were ground in an agate pestle and mortar, then mixed with an approximately equal volume of silicon powder and ground together to obtain a uniform powder.  This mixture was loaded into a 0.3 mm glass capillary.  The silicon serves primarily to disperse the \tto{} in the beam, while minimizing absorption, but also provides a convenient calibrant for wavelength and lattice parameters ($a_\mathrm{Si}=5.431194$ \AA, NIST powder diffraction standard 640c).  We measured the diffraction pattern of the mixture at room temperature, using the high resolution powder diffractometer of the Materials Science Beamline at the Swiss Light Source.  The diffractometer operates in Debye-Scherrer geometry, using a Mythen microstrip detector and capillary spinner. The incident wavelength was $0.620474(3)$ \AA~(i.e. 19.98 keV) with the $2\theta$ range extending from $2^\mathrm{\circ}$ to 120$^\mathrm{\circ}$.  Discernible peaks were visible to the highest scattering angles, and accounting for typical peak widths, there are around 270 effectively independent reflections (i.e. separated by $1.0\times FWHM$).

The work of Ross {\it et. al.} on Yb$_2$Ti$_2$O$_7$ suggests that rare earth pyrochlore crystals may be susceptible to small levels of disorder, in which some titanium ions are lost and are replaced during crystal growth by rare earth ions, and that this may be diagnosed by accurate measurements of the lattice parameter \cite{Ross:2012dj}.  In the work of Taniguchi {\it et al.}~\cite{taniguchi}, small variations of \tto{} of the type Tb$_{2+x}$Ti$_{2-x}$O$_y$ were studied, and the dependence of the lattice parameter on $x$ was reported.  From our x-ray diffraction measurements and the profile refinements discussed below, we find that the lattice parameter of our sample is $10.155288(1)$, which implies a stoichiometry of Tb$_{2.013}$Ti$_{1.987}$O$_{6.99}$, when compared to the reported lattice parameter trend.  In Ref.~\cite{taniguchi} it is reported that $x$ may vary by $\pm 0.002$ from the nominal value, which forms the main limitation for our determination of $x$ in this way.  Literature values of the lattice parameter of \tto{} are clustered around 10.154 \AA, but outlying values do exist and could cast doubt on this assignment of composition~\cite{Subramanian:1983vd,Han:2004bz,Lau:2006uc,Rule:2009wa}.

The powder diffraction data were modeled and fitted using the Rietveld method, as implemented in the package FullProf~\cite{fullprof}.  Preliminary refinements were performed in which the absorption coefficient was estimated from the quantitative phase analysis and updated between cycles of refinement of all other parameters, until the phase proportions were stable.  The data were subsequently corrected for absorption by the sample and capillary, and the background contribution from an empty capillary was subtracted.  These corrected data were used for the final refinements.

Using the Lebail technique, a structureless profile match was performed.  The profile matching showed that the shape of the Bragg peaks due to the silicon is well modeled by a pseudo-Voigt form, but that the Bragg peaks of \tto{} are best described by a pure Lorentzian.  Asymmetry effects are only important for $2\theta<11.5^\mathrm{\circ}$, and were refined independently for the two phases.  Subsequently we used a conventional Rietveld refinement of a crystallographic model incorporating two phases (i.e. \tto{} and silicon).  In general, we refined two capillary offset corrections, linearly interpolated background, profile parameters, and thermal parameters for both phases.   We found that anisotropic thermal parameters lead to some, or all, thermal ellipsoids becoming non-positive definite, and no marked increase in fit quality, so we retained isotropic thermal parameters throughout.   For \tto{}, we also refined the free positional parameter of the $48f$ oxygen site, and the lattice parameter.    The lattice parameter of silicon was held fixed; the wavelength and zero-shift of $2\theta$ were derived from an independent refinement of the silicon standard; and the asymmetry parameters from the profile matching were used.  All models were then refined freely to convergence.

\begin{figure*}
\includegraphics[scale=0.4,trim=0 20 0 20,angle=0]{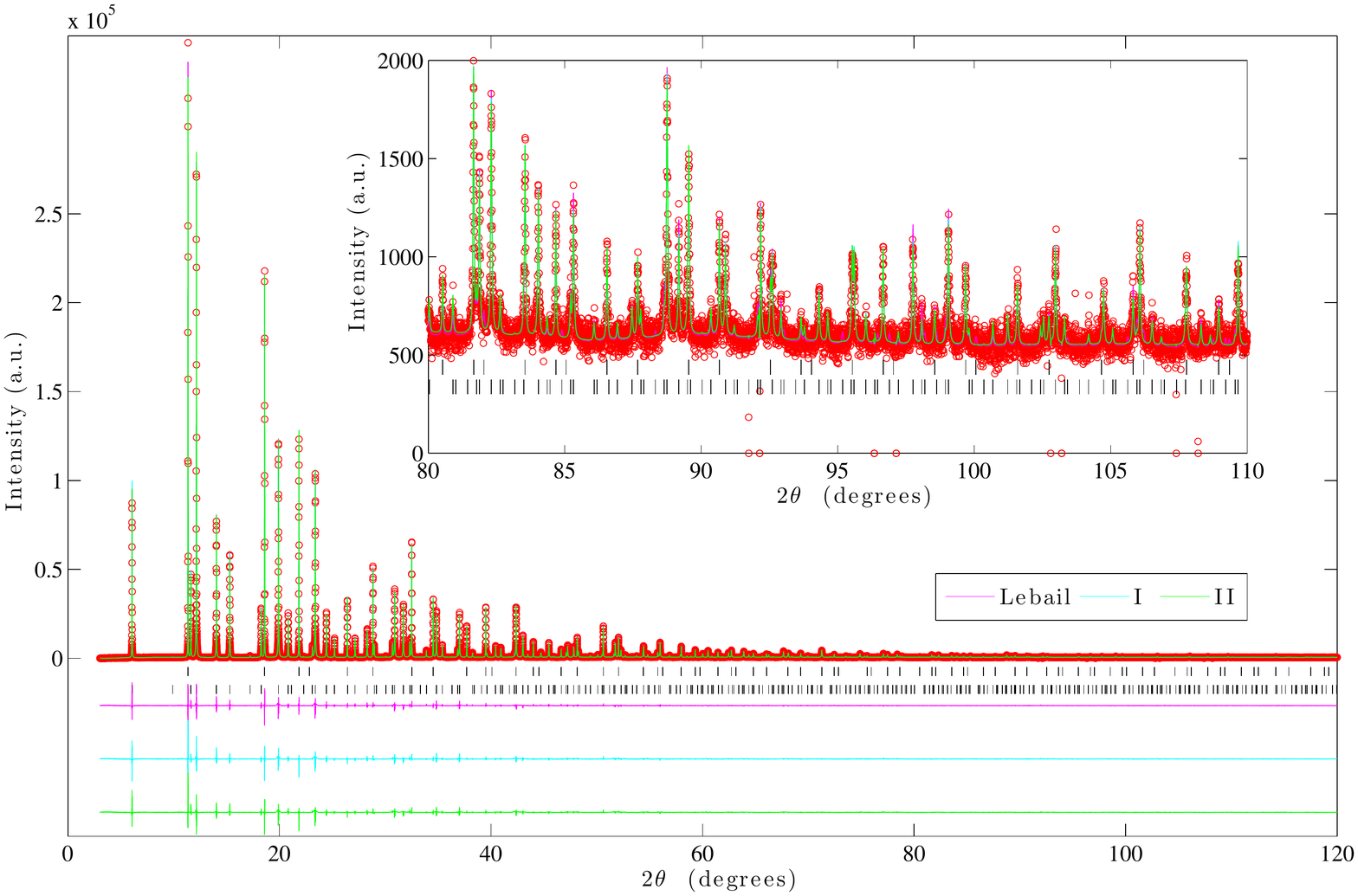}
\caption{Refinement of models $\mathrm{I}$ and $\mathrm{II}$ against powder x-ray diffraction data.  The similarity in quality of the crystallographic refinements and Lebail extraction implies that the fit cannot be significantly improved.}
\label{refinefig}
\end{figure*}

We have compared a model of stoichiometric \tto{} (model $\mathrm{I}$) with one with refinable occupancy of the $16d$ sites, Tb$_2$Ti$_{2-x}$Tb$_x$O$_{6-x/2}$O (with oxygen loss only from the $48f$ sites) (model $\mathrm{II}$).  The atom positions are summarized in Table~\ref{xtaltable}.  We found that the we do not have sufficient sensitivity to the defect population to provide any improvement on the estimate derived from the lattice parameter.  We illustrate the refinements in Supporting Fig.~\ref{refinefig}, and summarize the refinements in Table ~\ref{atomstable}.

We also investigated the crystal structure by single crystal neutron diffraction, which may have enhanced contrast for lighter elements, and also allowed us to compare the structure at room temperature, and in the magnetoelastic regime.  We used the TRICS diffractometer at the PSI, and a small sample of size $3\times3\times0.5$ mm cut from the main crystal.  We found that our refinements were dominated by extinction corrections, and that models with small levels of defects consistent with the (x-ray) lattice parameter could not be distinguished statistically from the stoichometric model.  We also found that the cubic symmetry is unmodified between room temperature and 5 K, so no structural change accompanies the onset of the magnetoelastic spin liquid at $T\sim 20$ K.  The periodicity of the excitations measured at low temperatures further suggests that no distortion occurs between 5 K and 0.05 K.

{\it v: Discussion and Summary.} By virtue of the lattice parameter determination, our powder x-ray diffraction measurements suggest that there is a degree of replacement of Ti$^{4+}$ by Tb$^{3+}$ at the level of $0.7\pm0.1$ \%, or Tb$_{2.013\pm0.002}$Ti$_{1.987\pm0.002}$O$_{6.9935\pm0.001}$.  Neither our x-ray or neutron diffraction measurements are able distinguish the defect level more precisely in a crystallographic structure refinement, but are also consistent with this level of defects.  The defect population is lower than that found by Ross {\it et al.} when comparing stoichiometric powders and crystals of Yb$_2$Ti$_2$O$_7$ (2.3\%)~\cite{Ross:2012dj}.  It is just outside the window of compositions studied by Taniguchi {\it et al}~\cite{taniguchi}, and they do not report heat capacity measurements for $x>0.005$, concluding instead that the spin liquid occurs only for $x<x_c$ and long range order occurs for all $x>x_c$ ($x_c=-0.0025$).  This picture is not completely compatible with this work, which suggests that the spin liquid typically studied in \tto{} single crystals by neutron scattering is a property of ``lightly stuffed'' samples with additional rare earth ions.  The study of ultrapure powders does not extend far to the ``stuffed'' side of ideality, even to where our sample lies, so we suggest that there is a narrow dome of stability for the ordered phase observed in Ref.~\cite{taniguchi}, with disordered phases on either side.  It would be very interesting to establish if the disordered phases are the same.  We expect that the MEM detailed here will exist in all other normal single crystal samples so far studied, and that they are to be found on the ``lightly stuffed'' side of the ideal stoichiometry.  Comparisons of crystallographic data at room temperature and 5 K, specific heat data in the range 0.35-50 K, and the periodicity of the excitations at 0.05 K show no evidence of a structural phase transition throughout the temperature range we study.

\begin{table*}
\begin{center}
\begin{tabular}{cccccc}
\hline
\hline
Atom & Wykoff Position  & x & y & z & Ideal Occupancy \\
\hline
Tb & $16c$ & 0 & 0 & 0 & 0.08333 \\
Ti & $16d$ & 0.5 & 0.5 & 0.5 & 0.08333 \\
O & $48f$ & $x$ & 0.125 & 0.125 & 0.25000 \\
O & $8a$ & 0.125 & 0.125 & 0.125 & 0.04167 \\
\hline
\hline
\end{tabular}
\end{center}
\caption{Atom positions of the spacegroup $Fd\bar{3}m$ used in the structural refinements.  These correspond to the second origin setting given in the International Tables of Crystallography. Nominally $x=0.42$, and occupancies are normalized to multiplicity of a general site (192).}
\label{xtaltable}
\end{table*}

\begin{table*}
\begin{center}
\begin{tabular}{cccccc}
\hline
Model $\mathrm{I}$ &(52 variables) & & & &\\
$R_p$ (\%) & 6.54&$R_{wp}$ (\%) &9.26 &$R_{exp}$ (\%) &5.26\\
$R_\mathrm{Bragg, Tb_2Ti_2O_7}$ &3.11 & & & &\\
$\chi^2$ &3.1 & & & &\\
 $a$&10.15529(1) & & & &\\
\hline
Atom & $x$ & $y$ & $z$ & $B_{iso}$ & Occupancy\\
Tb &    0 & 0 & 0                  &0.700(3) & 0.08333    \\
Ti  &    0.5 & 0.5 & 0.5                  & 0.568(7)& 0.08333   \\
O$_{48f}$  & 0.42203(20)& 0.125 & 0.125& 0.850(41)&0.25       \\
O$_{8a}$   &  0.125&0.125&0.125 & 1.076(76)& 0.04167  \\
\hline
\hline
Model $\mathrm{II}$ &(53 variables)& & & &\\
$R_p$ (\%) & 6.6&$R_{wp}$ (\%) &9.18 &$R_{exp}$ (\%) &5.26\\
$R_\mathrm{Bragg, Tb_2Ti_2O_7}$ &3.34 &  & & \\
 $\chi^2$ &3.05 & & & &\\
 $a$ &10.15529(1) & & & &\\
\hline
Atom & $x$ & $y$ & $z$ & $B_{iso}$ & Occupancy\\
Tb &    0 & 0 & 0                  &0.685(3) & 0.08333    \\
Ti  &    0.5 & 0.5 & 0.5                  & 0.731(10)& 0.081(0)   \\
Tb & 0.5 & 0.5 & 0.5 & 0.731(10) & 0.002(0) \\
O$_{48f}$  & 0.42129(20)& 0.125 & 0.125& 0.757(40)&0.249(0)       \\
O$_{8a}$   &  0.125&0.125&0.125 & 1.132(77)& 0.04167  \\
\hline
\end{tabular}
\end{center}
\caption{Parameters for models $\mathrm{I}$ and $\mathrm{II}$ when refined against the powder x-ray diffraction data.}
\label{atomstable}
\end{table*}

\end{document}